\documentclass[twocolumn]{aa}
\usepackage{graphicx}
\usepackage{txfonts}
\usepackage{enumerate}

\begin{document}

\title{Particle acceleration in thick parallel shocks with high
       compression ratio}

\headnote{Research Note}

\author{Joni J. P. Virtanen \inst{1} \and  Rami Vainio \inst{2}}
\offprints{J.Virtanen}

\institute{%
  Tuorla Observatory, V\"ais\"al\"a Institute for Space Physics and Astronomy,
  V\"ais\"al\"antie 20, FI-21500 Piikki\"o, Finland \\
  \email{joni.virtanen@utu.fi}
  \and
  Department of Physical
  Sciences, P. O. Box 64, FI-00014 University of Helsinki, Finland \\
  \email{rami.vainio@helsinki.fi}
}

\date{Received 2.5.2005 / Accepted 1.6.2005}

\abstract{We report studies on first-order Fermi acceleration in
  parallel modified shock waves with a large scattering center
  compression ratio expected from turbulence transmission
  models. Using a Monte Carlo technique we have modeled particle
  acceleration in shocks with a velocity ranging from
  nonrelativistic to ultrarelativistic and a thickness extending
  from nearly steplike to very wide structures exceeding the
  particle diffusion length by orders of magnitude. The
  nonrelativistic diffusion approximation is found to be
  surprisingly accurate in predicting the spectral index of a thick
  shock with large compression ratio even in the cases involving
  relativistic shock speeds.   }

\authorrunning{Virtanen \& Vainio}
\titlerunning{Particle acceleration in thick shocks with high compression}
\maketitle

%
%
%
%

\section{Introduction}

First-order Fermi acceleration in shocks with steplike velocity profile
is well known to be an efficient way of producing nonthermal
particle populations with power law %
(see, e.g., reviews of Drury (\cite{Drury1983}) and Kirk \& Duffy 
(\cite{KD1999}) for nonrelativistic and relativistic studies, respectively).
In the nonrelativistic regime with infinitely thin step-shocks this 
mechanism produces %
particle momentum distributions $f(p) \propto p^{-s}$ with
spectral index $s$ depending only on the compression ratio
of the shock, $r$, as \( s= 3r/(r-1) \).
For the relativistic case also the shock speed $V_1$
affects, and the spectral index -- in the case of isotropic 
pitch-angle diffusion -- can be written as
\begin{equation}
  s_{KW} = \frac{3 V_1 c^2 - 2 V_1 V_2^2 + V_2^3}{(V_1-V_2) c^2}
  = \frac{3r}{r-1}\left(1 - \frac{V_1^2(2r-1)}{3 r^2 c^2}\right)
  \label{eq:s_KW}
\end{equation}
as was recently shown by Keshet \& Waxman (\cite{KeshetWaxman2005}).

In a modified shock -- i.e., in a shock with a nontrivial velocity
profile and finite shock thickness -- the acceleration efficiency
drops as the thickness increases (e.g., Schneider \& Kirk
\cite{SK1989}, Virtanen \& Vainio \cite{VV2003a}). Again, in the
nonrelativistic case the resulting spectral index as a function of
shock thickness and compression ratio can have a simple analytical
solution, found by Drury et al.\ (\cite{DruryEtAl1982}), whereas the
matter becomes much more complicated in the relativistic regime (e.g.,
Schneider \& Kirk \cite{SK1989}).  In their nonrelativistic study
Drury et al. (\cite{DruryEtAl1982}) showed that while the produced
spectral index tends to the well known step-shock limit when the shock
thickness approaches zero, the high energy part of the spectrum can
have a power law even when the transition is large compared to free
path of an accelerating particle. The produced spectral index depends
on the thickness of the transition region and the compression ratio as
\begin{equation}
  s  = \frac{3V_{1}}{V_{1}-V_{2}}\left(1+\frac{1}{\beta}
    \frac{V_{2}}{V_{1}-V_{2}}\right)
  =  \frac{3r}{r-1}\left(1+\frac{1}{\beta(r-1)}\right),
  \label{eq:s_drury}
\end{equation}
where %
$\beta$ is a parameter inversely proportional to the
shock thickness.  For a self-consistent cosmic-ray dominated shock it
has the value of $\beta=\frac 12 (1+\gamma_c)$, where $\gamma_c$ is
the adiabatic index of the cosmic-ray gas. Later Schneider \& Kirk
(\cite{SK1989}) found the nonrelativistic diffusion approximation with
spectral index of the form of Eq.\ (\ref{eq:s_drury}) to agree well
even in cases in which the thickness of the shock transition is of the
order of the particle mean free path where the diffusion approximation
should not be mathematically justified. In the relativistic regime the
diffusion approximation was, expectedly, found to fail to approximate
the produced spectral index. %
One should also note that Drury et al.\ (\cite{DruryEtAl1982}) and Schneider
\& Kirk (\cite{SK1989}) neglected the second-order Fermi acceleration in their
studies. The effect of this mechanism in the shock downstream has been studied
analytically in modified nonrelativistic shocks by Schlickeiser
(\cite{Schlickeiser1989}) and numerically in relativistic shocks by Virtanen
\& Vainio (\cite{VV2005}), and it leads to more complex dependence of the
produced spectral index of the shock's Peclet and Alfv\'enic Mach number and
compression ratio.

From Eq.\ (\ref{eq:s_drury}) it becomes obvious that even shocks with
large thickness (i.e., small $\beta$) can accelerate particles
efficiently if only the compression ratio is sufficiently high. This
may seem irrelevant, as the gas compression ratio of a shock is
generally limited to rather small values, e.g.,
$r<(\gamma_c+1)/(\gamma_c-1)$ for cosmic-ray dominated
nonrelativistic shocks. However, the compression ratio felt by the
accelerated particles is not that of the flow; as already noted by
Bell (\cite{Bell1978}), one needs to consider the finite phase speed
of the hydromagnetic waves responsible for particle scattering around
the shock waves. Taking the fluctuations to be Alfv\'en waves, the
problem of turbulence transmission at the shock is tractable and the %
effective scattering center compression ratio is found to be much
larger than the gas compression ratio for shocks of low Alfv\'enic
Mach number (Vainio \& Schlickeiser \cite{VS1998,VS1999}; Vainio et al.\
\cite{VVS2003,VVS2005}).

%
%
%
%

\section{Scattering-center compression ratio}

\begin{figure}
\includegraphics[width=1.00\linewidth]{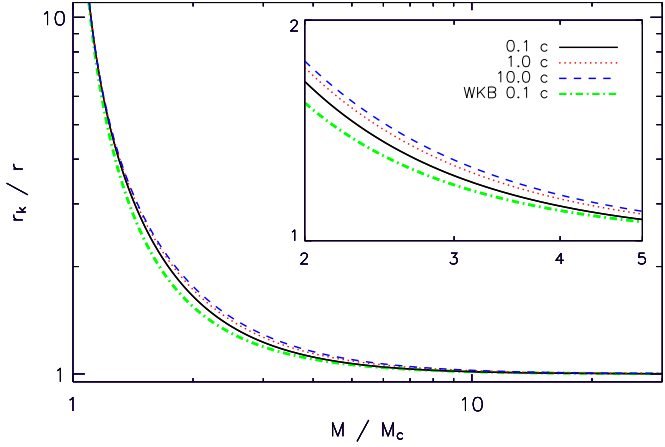}
\caption{Ratio of the scattering center compression ratio $r_{k}$ to
  that of the gas $r$ for different Alv\'enic Mach numbers $M$. The
  spectral index of the turbulence power spectrum is $q=2$, corresponding 
  to a momentum-independent particle mean free path. $M_{\rm c}=\sqrt{r}$.
  See text for details.}
\label{fig:rmu}
\end{figure}

As shown by Vainio \& Schlickeiser (\cite{VS1998}) for nonrelativistic
and Vainio et al.\ (\cite{VVS2003,VVS2005}) for relativistic speeds,
shocks with low to moderate quasi-Newtonian Alv\'enic Mach number
$M\gtrsim\sqrt{r}$ transmit the Alfv\'en waves from upstream to
downstream so that most of the downstream waves propagate
antiparallel to the direction of the shock-frame plasma flow. Thus,
the scattering center compression ratio, calculated assuming a
vanishing upstream cross-helicity,
\begin{equation}
  r_{k}=\frac{V_1(1+H_{\rm c2}V_{\rm A2}V_2/c^2)}{V_2+H_{\rm c2}V_{\rm A2}}
\label{eq:r_k}
\end{equation}
tends to infinity as the Alfv\'enic Mach number of the shock
$M=u_{1}/u_{{\rm A}1}$ (where $u_{1}$ and $u_{{\rm A}1}$ are the shock
proper speed and the upstream proper Alfv\'en speed respectively, both
measured in the rest frame of the upstream plasma) approaches the
critical Mach number $M_{{\rm c}}=\sqrt{r}$, since $V_2\to V_{\rm A2}$
and $H_{\rm c2}\to-1$ at this limit. Here, $V_{{\rm A2}}$ and $H_{\rm
c2}$ are the Alfv\'en speed and the wave cross helicity in the
downstream region, and they can be calculated from the shock jump
conditions. This is shown in Fig.~\ref{fig:rmu} for parallel shocks
with three velocities assuming a momentum independent scattering mean
free path and a vanishing upstream cross-helicity. For details on how
to compute $r_k$, see Vainio et al.\ (\cite{VVS2003,VVS2005}).

The scattering hydromagnetic fluctuations have wavelengths comparable
to the Larmor radius of the resonant particles. Thus, for shock waves
discussed in this paper, the Alfv\'en waves actually see the shock
front as a thick structure. Thus, their transmission becomes governed
by the WKB theory. A detailed analysis of this case is beyond the
scope of this short Note but qualitatively one expects wave
transmission in these shocks to show very similar behavior as in
steplike shock waves: antiparallel propagating waves are amplified
much more than the parallel propagating ones because of the
conservation of the wave action, $(V\pm V_{\rm A})^2P^\pm(f,x)/V_{\rm
A} = \mbox{const.}$, where $f=(V\pm V_{\rm A})k/2\pi$ and
$P^{+(-)}(f,x)$ are the (conserved) wave frequency and the power
spectrum of the parallel (antiparallel) propagating waves,
respectively. For the nonrelativistic case and for spectral index $q$
of the fluctuations one finds after a short calculation
\begin{equation}
  I_2^\pm(k)=I_1^\pm(k) r^{q+1/2}\frac{(M\pm 1)^{q+1}}{(M\pm
  r^{1/2})^{q+1}}
\end{equation}
for the downstream intensities of waves propagating
parallel (antiparallel) to the flow, $I^{+(-)}_{1[2]}$. For a
vanishing upstream cross helicity we get
\begin{equation}
  H_{\rm c2} = \frac{(M+ 1)^{q+1}(M-\sqrt{r})^{q+1}
    -(M-1)^{q+1}(M+\sqrt{r})^{q+1}}
  {(M+ 1)^{q+1}(M-\sqrt{r})^{q+1}
    +(M-1)^{q+1}(M+\sqrt{r})^{q+1}}.
\end{equation}
Thus, $r_{k}\to \infty$ as $M\to \sqrt{r}$, as in our earlier
calculations. An example of this is shown in Fig.~\ref{fig:rmu} for
nonrelativistic shock (dash-dot-dashed line, labeled ''WKB $0.1\,c$''). 
Of course, the singularity is not reached in practice
because the wave pressure in the downstream region becomes large and
makes the compression ratio stay below $M^2$ in a self-consistent
calculation (Vainio \& Schlickeiser \cite{VS1999}). Nevertheless,
large scattering compression ratios are expected for low-Mach number
shocks even for large shock thickness.

%
%
%
%

\section{Particle acceleration}

We have used test-particle Monte Carlo simulations to calculate the %
particle energy distribution \( N(E) \propto E^{-\sigma} \)
spectral indices \( \sigma = s - 2 \)
resulting from first-order Fermi acceleration at the
shock for parallel shocks. Detailed description of the code is given
elsewhere (Appendix A of Virtanen \& Vainio \cite{VV2005}). The
simulations follow test-particles in a one dimensional flow.  Small
pitch angle scatterings are performed in the frame comoving with the
scattering centers. Instead of using two scattering wave fields,
$I^\pm$, we calculate the effective wave speed as $H_{\rm c}V_{\rm A}$
and apply only a single scattering wave field propagating at this
speed relative to the plasma. This eliminates stochastic acceleration
from the model and allows us to study the effect of first-order Fermi
acceleration at the shock, only.

We use the hyperbolic tangent function of Drury et al.\ %
(\cite{DruryEtAl1982}) for the scattering center speed (denoted
hereafter by $V$) profile across the shock:
\begin{equation}
  V(x)=V_{1}-\frac{V_{1}-V_{2}}{2}\left[1+
    \tanh\left(\frac{3\beta(V_{1}-V_{2})}{2c}\frac{x}{\lambda}\right)\right].
  \label{eq:flowprofile}
\end{equation}
Here, $\lambda$ is the (constant) particle mean free path and the
parameter $\beta$ the same as in Eq.\ (\ref{eq:s_drury}). In this
study we consider $\beta$ as a free parameter and simulate shocks of
different widths by varying its value. Defining the diffusion length
of the particles in the usual nonrelativistic manner as $d_i=\frac 13
\lambda c/V_i$, we can write the inverse shock thickness as
$W^{-1}=\frac12\beta(d_1^{-1}-d_2^{-1})$. Thus, nonrelativistic
simulations with $\beta\ll 1$ correspond to shocks with a thickness
much larger than the diffusion length of the accelerated particles.

For each simulation we fix the proper speed of the shock relative to
the upstream scattering centers,
\mbox{\(u_{1}=\Gamma_{1}V_{1}=c(\Gamma_{1}^{2}-1)^{1/2}\)}, and the
(scattering center) compression ratio $r_k=V_{1}/V_{2}$. Particles are
given a small initial energy and they are injected into the shock in
the upstream so that they have time to isotropize before reaching the
shock transition region.
Particles are then traced in the shock until they hit the downstream
escape boundary at $x_{2}$. This boundary is set sufficiently far away
from the shock so that the particles have enough time to isotropize in
the rest frame of the downstream scattering centers. A probability of
return is calculated and applied for each particle crossing the boundary;
the method is described e.g., by Ellison et al.\ %
(\cite{EllisonEtAl1990}). Particle splitting is used to improve the
statistics at high energies.
Energy losses due to synchrotron emission are not considered in the
simulations.

%
%
%
%

\begin{figure}
  \includegraphics[width=\linewidth]{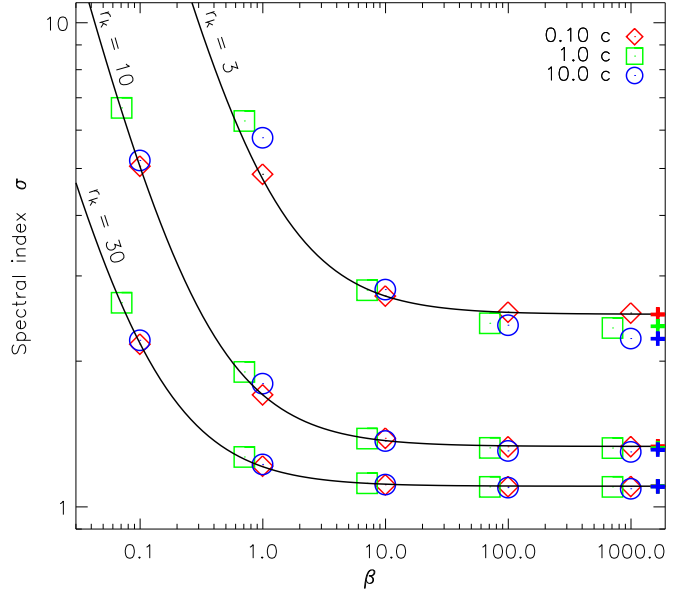}
  \caption{%
    Particle energy
    spectral indices as a function of $\beta$ for shock proper
    speeds $u_{1}=0.1\, c$ (diamonds), $1.0\, c$ (squares), and
    $10.0\, c$ (circles) for scattering center compression ratios
    $r_k=3$, $10,$ and $30$; 
    the crosses at the right end denote values obtained from 
    Eq.\ (\ref{eq:s_KW}) for step shocks ($\beta \to \infty$).
    Solid lines show the predictions of the
    nonrelativistic diffusion approximation.  The statistical errors
    of the fitted spectral indices are of the order of $0.040$ or less
    for all datapoints.  }
  \label{fig:results}
\end{figure}

Simulations were run for shock proper speeds $u_{1}=\{0.1,1,10\}\,
c$, compression ratios $r_k=\{3,10,30\}$, and for $\beta$ having
values between $\sim\,$$0.07$ and $\sim\,$$1000$.
The resulting spectral indices are shown in Fig.\ \ref{fig:results}
together with the predictions of the nonrelativistic diffusion
approximation from Eq.\ (\ref{eq:s_drury}).

In the nonrelativistic case ($V_1\approx u_1 = 0.1\, c$) the results
are as expected: spectral indices are in accordance with the diffusion
approximation in all cases, even where the shock thickness is
comparable to the diffusion length of the particles.  
This is the case with %
     all compression
ratios studied. 
For the mildly relativistic case ($u_1 = 1.0\, c$, i.e., $V_1 \approx
0.7\, c$ and $\Gamma_1 \approx 1.4 $) and the lowest compression ratio
$r_k=3$ (corresponding to the gas compression ratio of a relativistic
parallel shock) the simulated indices start to differ from the
diffusion approximation, and the differences grow even larger for the
ultrarelativistic case ($u_1 = 10\, c$ with $V_1 \approx 0.995\, c$
and $\Gamma_1 \approx 10 $). At the step-shock limit, where $\beta \to
\infty$, the index from the ultrarelativistic shock tends to 
 $\sigma \approx 2.2$
 as expected, and also the $u_1 = c$ case produces indices
harder than the diffusion theory predicts.  This, of course, is
expected from previous studies.  As the shock thickness increases
($\beta$ decreases) behavior similar to that reported by Schneider \&
Kirk (\cite{SK1989}) is seen: while for thin shocks the diffusion
approximation predicts harder spectra than those obtained from
simulation, for shocks with wider transition the spectra are softer
than the values from the diffusion approximation.

The most interesting result of our study is, however, that as the
compression ratio increases above the traditional gas-compression
ratio values, the nonrelativistic diffusion approximation works
rather well even for relativistic shocks. Small deviations are seen in
all cases where relativistic effects are present, but the spectral
indices can be approximated by Eq.\ (\ref{eq:s_drury}) to an accuracy
of the relevant observations, especially for the case
$r_k=30$. Furthermore, even shocks with very wide transition can
produce spectra that are hard, compared to those usually considered to
be produced in parallel shocks.

\section{Conclusions}

Based on our analysis, we arrive at the following conclusions of
particle acceleration at modified shocks.
\begin{enumerate}[I.]
\item Alfv\'en-wave transmission can lead to large scattering center
  compression ratios in parallel shocks regardless of the thickness of
  the shock front if the Alfv\'enic Mach number of the shock is close
  to the critical value \mbox{$M_{\rm c}=\sqrt{r}$.}
\item Modified parallel shocks can accelerate particles efficiently
  even if their thicknesses exceed the diffusion length of the
  accelerated particles, if the scattering center compression ratio is
  large.
\item Nonrelativistic theory of particle acceleration in modified
  shocks, Eq.\ (\ref{eq:s_drury}), gives a  good
  approximation to the spectral index at shocks with large
  scattering center compression ratio even for ultrarelativistic
  shock velocities.
\end{enumerate}

%
%
%
%

\begin{acknowledgements}
  CSC, the Finnish IT center for science, is acknowledged for providing 
  part of the computing facilities.
\end{acknowledgements}

%
%
%
%

\end{document}